\documentclass{emulateapj}
\usepackage{apjfonts}
\usepackage{html}

\def\gsim{\mathrel{\rlap{\lower2pt\hbox{\hskip0pt\small$\sim$}}
   \raise2pt\hbox{\small $>$}}}    
\def\lsim{\mathrel{\rlap{\lower2pt\hbox{\hskip0pt\small$\sim$}}
   \raise2pt\hbox{\small $<$}}}    
\def\beq{\begin{equation}} 
\def\eeq{\end{equation}} 

\shorttitle{PRE-REIONIZATION 21-CM STRUCTURE}
\shortauthors{ALVAREZ, PEN \& CHANG}

\submitted{Submitted to ApJ Letters June 30, 2010}

\begin{document}

\title{Enhanced detectability of pre-reionization 21-cm structure} 

\author{Marcelo A. Alvarez\altaffilmark{1}, Ue-Li Pen\altaffilmark{1}, and Tzu-Ching Chang\altaffilmark{1,2}}

\altaffiltext{1}{Canadian Institute for Theoretical Astrophysics,
 University of Toronto, 60 St.George St., Toronto, ON M5S 3H8,
 Canada} 
\altaffiltext{2}{IAA, Academia Sinica, PO Box 23-141, Taipei, 10617, Taiwan} 
\email{Electronic address: malvarez@cita.utoronto.ca}

\begin{abstract}
Before the universe was reionized, it was likely that the spin
temperature of intergalactic hydrogen was decoupled from the CMB by UV
radiation from the first stars through the Wouthuysen-Field effect.
If the IGM had not yet been heated above the CMB temperature by that
time, then the gas would appear in absorption relative to the CMB.
Large, rare sources of X-rays could inject sufficient heat into the
neutral IGM, so that $\delta T_b~>~0$ at comoving distances of tens to
hundreds of Mpc, resulting in large 21-cm fluctuations with $\delta T_b\simeq
$~250~mK on arcminute to degree angular scales, an order of magnitude larger in amplitude than that
caused by ionized bubbles during reionization, $\delta T_b\simeq
25$~mK.   This signal could therefore be easier to detect and probe
higher redshifts than that due to patchy reionization.  For the case
in which the first objects to heat the IGM are QSOs hosting
10$^7-M_\odot$ black holes with an abundance exceeding $\sim 1$~Gpc$^{-3}$ at $z\sim 15$, 
observations with either the Arecibo Observatory or the Five Hundred Meter 
Aperture Spherical Telescope (FAST) could detect and image their
fluctuations at greater than 5-$\sigma$ significance in about a month of
dedicated survey time.  Additionally, existing facilities such as MWA
and LOFAR could detect the statistical fluctuations arising from a
population of $10^5-M_\odot$ black holes with an abundance of $\sim
10^4$~Gpc$^{-3}$ at $z\simeq 10-12$.

\end{abstract}


\section{Introduction}

The 21 cm transition is among the most promising probes of the
high-redshift universe.   The observed differential brightness
temperature from the fully neutral IGM at the mean density with spin temperature
$T_{\rm s}(z)$ is given by
\beq
\delta T_b(z) \simeq 29~{\rm mK}
\left(\frac{1+z}{10}\right)^{1/2}
\left[1-\frac{T_{\rm CMB}(z)}{T_{\rm s}(z)}\right]
\eeq
\citep[e.g.,][]{madau/etal:1997}.
In the absence of an external radiation field, only the hot, dense gas in
minihalos would have been able to develop a spin temperature different
from that of the CMB at $z\sim 20$ 
\citep[][]{shapiro/etal:2006}.  However, radiation emitted by
the earliest generations of stars
\citep[e.g.,][]{ciardi/madau:2003,barkana/loeb:2005} and X-ray 
sources \citep[][]{chuzhoy/etal:2006,chen/miralda-escude:2008}  
could have coupled the spin temperature to the IGM temperature via the
``Wouthuysen-Field effect'' \citep{wouthuysen:1952,field:1959} long
before reionizing the universe, at redshifts as high as $z=20-30$,
causing the IGM to appear in absorption with respect to the CMB, with
$\delta T_b\simeq -250$~mK. 

Only a modest amount of heating is necessary to raise the gas
temperature above the CMB temperature of $\sim 30 - 60$~K at $z\sim 10
- 20$.  By the time reionization was well underway, it is generally
believed that the neutral component of the IGM had already been heated to $T_{\rm
  s}\gg T_{\rm CMB}$, and therefore the observed brightness
temperature will be an order of magnitude lower in amplitude, $\delta
T_b\simeq 25$~mK, than when the IGM was in absorption
\citep[e.g.,][]{furlanetto/etal:2006}.  

Clearly, there should be some transition epoch, during which sources
of X-ray radiation heated the neutral IGM, creating ``holes'' in the
absorption.  Natural candidates for these early sources of heating are quasars
\citep[e.g.,][]{chuzhoy/etal:2006,zaroubi/etal:2007,thomas/zaroubi:2008,chen/miralda-escude:2008}.
Unfortunately, little is known about the quasar population at $z>6$,
with the only constraints coming from the bright end at $z\sim 6$
\citep[e.g.,][]{fan/etal:2002,willott/etal:2010}, with
inferred black hole masses $\gsim 10^8~M_\odot$, luminosities $\gsim
10^{46}$~erg~$s^{-1}$, and a comoving abundance $\gsim 1$~Gpc$^{-3}$.      

Quasars are not the only high-redshift objects able to produce X-rays.
High-redshift supernovae \citep{oh:2001} and X-ray binaries could have
also been sources. Theoretical models typically parametrize X-ray
production associated with star formation by $f_X$, normalized so that
$f_X=1$ corresponds to that  observed for local starburst galaxies
\cite[e.g.,][]{furlanetto:2006,pritchard/furlanetto:2007}.  

Our aim in this paper is to explore the enhanced 21-cm signature of early
quasars.  We will therefore assume that the 
first galaxies were efficient enough to couple the spin temperature to
the kinetic temperature by $z\simeq 20$, but did not
heat the IGM above the CMB temperature, i.e. $f_X\ll 1$.  
 As shown by \citet{pritchard/loeb:2010}, even for values of $f_X>1$ it
is still possible that there is a significant period between the time
when the spin temperature was coupled to the kinetic temperature and
the time when the mean
IGM was heated above the CMB temperature.  In addition, the first galaxies were
likely clustered around the quasars and may have actually increased
the sizes of the heated regions.  On the other hand, X-ray-emitting galaxies could
have also formed outside the regions that would have been heated by
the quasars, making the 21-cm features due to early, rare quasars that
we predict here less prominent. 

This paper is organized as follows.  In \S 2 we present our forecasted signal from
partially-heated regions and a brief discussion of the
expected black hole abundance at high-redshift.  In \S 3 we assess the detectability of
the predicted signal under various assumptions about the early quasar
population.  We end in \S 4 with a discussion of possible survey
strategies and requirements of detection on the high-redshift black
hole abundance.   All calculations were done
assuming a flat universe with
$(\Omega_mh^2,\Omega_bh^2,h,n_s,\sigma_8)=(0.133,0.0225,0.71,0.96,0.8)$,
consistent with WMAP 7-year data \citep{komatsu/etal:2010}.  All
distances are comoving.   

\pagebreak

\section{Forecast}

Our predictions will focus on accreting black holes with
masses greater than $\simeq 10^5 M_\odot$.  It is of course possible
that a more abundant population of black holes with lower masses
existed at these early times, left behind as remnants of early
generations of massive Pop III stars
\citep{heger/etal:2003}, acting as ``seeds'' for the observed
$z\sim 6$ supermassive black hole population
\citep[e.g.,][]{li/etal:2007}, and powering ``miniquasars''
\citep{madau/etal:2004}.  However, radiative feedback from the
progenitor stars \citep{johnson/bromm:2007} as well as the
accretion radiation itself \citep{alvarez/etal:2009a}, would
have likely substantially limited their growth and corresponding X-ray
emission at early times.  An alternative scenario for forming the
seeds is gaseous collapse to black holes with masses greater than
$\simeq 10^4 M_\odot$ in the first halos with virial temperature
greater than $\simeq 10^4$~K 
\citep[e.g.,][]{bromm/loeb:2003,begelman/etal:2006}.   Formation of
black holes by this mechanism may have been quite a rare
occurence \citep[e.g.,][]{dijkstra/etal:2008}.  It is this scenario, in
which most accreting black holes in the universe were relatively rare
and more massive than $\simeq 10^5 M_\odot$, that is most consistent
with the predictions we make here.  

Because the heating is a time-dependent
effect, we will parametrize the total energy radiated during
accretion as $E_{\rm tot}=Lt_{\rm QSO}=\epsilon M_{\rm BH}c^2$,  where
we take the radiative efficiency $\epsilon=0.1$.   In principle some
of the rest mass energy, i.e. the initial seed mass of the black
hole, did not contribute to heating the surroundings, but in general
the seed mass is expected to be small compared to the mass of the
black hole after it undergoes its first episode of
radiatively-efficient accretion as a quasar, so we neglect it.

\subsection{Quasar spectral energy distribution}
We assume the spectral energy distribution of the quasar is
given by a broken power-law, $S_\nu \propto \nu^{-0.5}$ at
$\nu<\nu_{\rm b}$, and $S_\nu \propto \nu^{-1.5}$ at
$\nu>\nu_{\rm b}$, with $h\nu_{\rm b}=11.8$~eV
\citep[e.g.,][]{bolton/haehnelt:2007b}, approximately 
consistent with the template spectra of \citet{telfer/etal:2002}.  If
the quasar has a total bolometric luminosity $L$, then the spectral
energy distribution is 
\beq 
S_\nu=\frac{L}{4\nu_{\rm b}}\left\{
\begin{array}{ll}
(\nu/\nu_{\rm b})^{-0.5}, & 
\nu<\nu_{\rm b},\\
(\nu/\nu_{\rm b})^{-1.5}, & 
\nu>\nu_{\rm b}.
\end{array}
\right.  
\eeq 

\subsection{The 21-cm profile around individual quasars} 

The temperature profile surrounding a quasar can be obtained by
considering the fraction of the radiated energy absorbed per atom at
comoving distance $r$ and redshift $z$,  
\beq
\Gamma_{\rm HI}(r,z)=\frac{(1+z)^2}{4\pi Lr^2}\int_{\nu_{\rm
    HI}}^{\infty}d\nu
\frac{S_\nu\sigma_\nu}{h\nu}
(h\nu-h\nu_{HI}) \chi_\nu \exp[-\tau_\nu(r)],
\eeq
where $\chi_\nu$ is the fraction of photoelectron energy,
$h\nu-h\nu_{\rm HI}$, which goes into heat, with the rest 
being lost to secondary ionizations and excitations.  In the limit in
which the ionized fraction of the gas is low, $0.1 < \chi_\nu < 0.3$
for all $h\nu>25$~ eV \citep{shull/vansteenberg:1985}.  In what
follows we will make the approximation that $S_\nu = 0.2$ for all
$\nu$, and assume that the energy radiated is $Lt_{\rm   qso}\equiv
\epsilon M_{\rm BH}c^2$, with $t_{\rm qso}$ short compared to the
Hubble time.  In this case, the relative brightness temperature is
given by 
\beq
\delta T_{\rm b}(M_{\rm BH},r,z)=29~{\rm mK}
\left[
1-\frac{T_{\rm CMB}(z)}{T_s(M_{\rm BH},r,z)}
\right]
\left[
\frac{1+z}{10}
\right]^{1/2},
\label{dtb}
\eeq
where
\beq
T_s(M_{\rm BH},r,z)=T_{\rm IGM}(z)+\frac{2\epsilon M_{\rm BH}c^2}{3k_b}
\Gamma_{\rm HI}(r,z).
\label{ts}
\eeq

\begin{figure}
\includegraphics[width=0.49\textwidth]{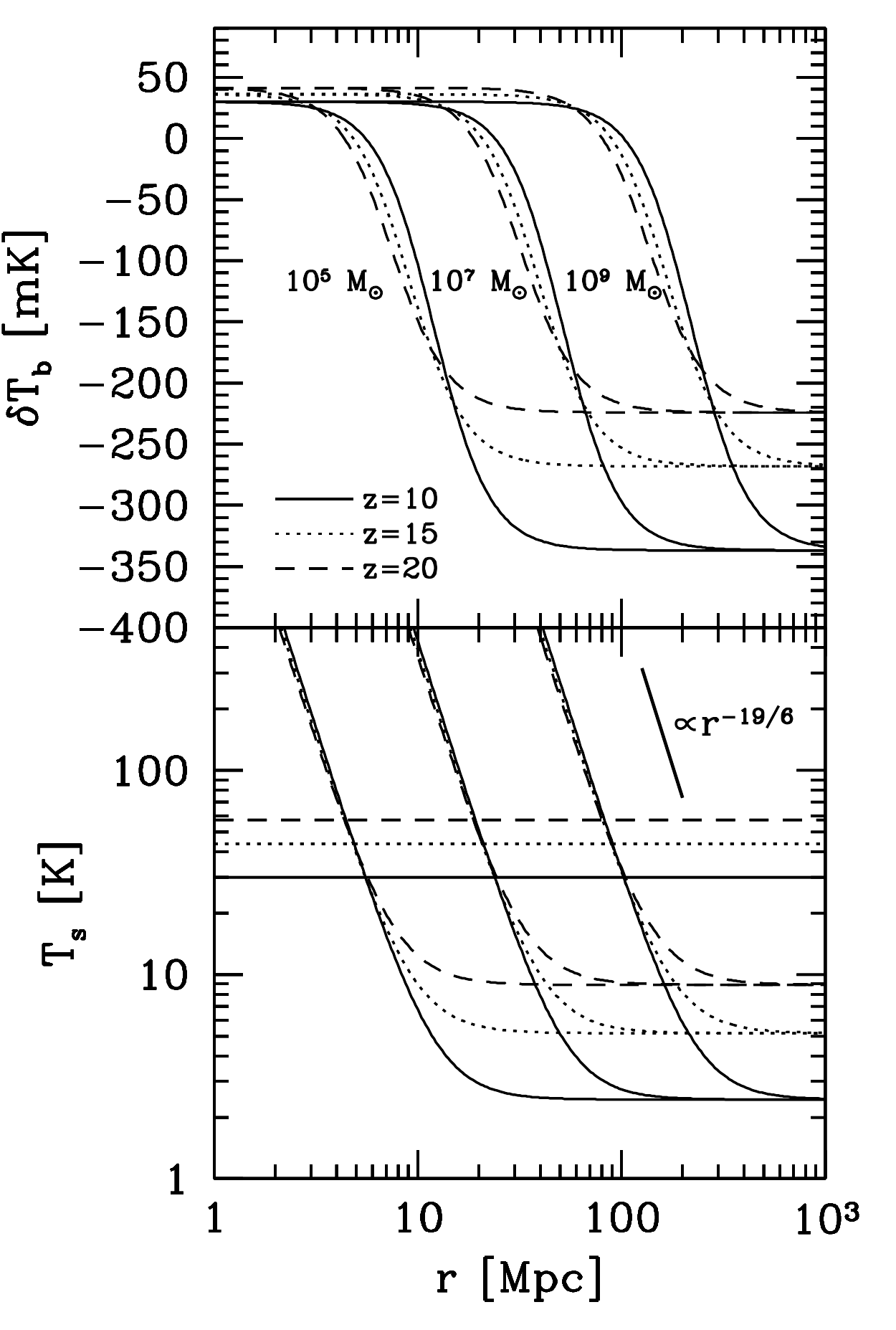}
\caption{  
Top: Profiles of the observed differential  brightness temperature versus
comoving distance at three different redshifts around a quasar that has
radiated 10 per cent of the rest-mass energy of $10^5$, $10^7$, and
$10^9$~$M_\odot$, as labeled. Bottom: Spin temperature of the IGM,
obtained via equation (\ref{ts}).
}
\label{fig1}
\end{figure}

 In calculating the optical depth, we assume that the IGM is
completely neutral at the mean density, $n_{\rm H}(z)$, so that
$\tau_\nu(r,z)=rn_{\rm H}(z)(1+z)^{-1}\sigma_\nu$.
In reality, the quasar's own H~II region will reduce the opacity at
small radii, but given that we are concerned with the small amount of
heating happening at much larger radii,  we neglect the H~II
region when calculating the optical depth.

Shown in Fig.~\ref{fig1} are profiles of the observed differential
brightness temperature versus comoving distance at three different
redshifts.  The total energy radiated by the quasar, $Lt_{\rm QSO}$ in each curve
corresponds to 10 per cent of the rest mass, as labeled.  Close to the
quasar, the gas is heated above the CMB temperature, and $\delta
T_b\simeq 30-40$~mK.  Further away, the heating from the quasar declines
due to spherical dilution and attenuation of the lowest energy
radiation, with $\delta T_b$ finally reaching about -220 to -340~mK at large
distances.   The FWHM size of the fluctuations are about 20, 80, and
400~Mpc for black holes of mass $10^5$, $10^7$, and $10^9$~$M_\odot$, respectively.

% \subsection{Poisson fluctuations}

% To estimate the fluctuating background from a superposition of sources, we
% assume all black holes at a given redshift have the same mass, $M_{\rm
%   BH}$,  and comoving number density, $n_{\rm BH}$.  We then
% generate a random realization of black hole positions within a
% periodic box, and calculate the heating from each source using
% equation (\ref{ts}).  

% Shown in Fig.~\ref{fig2} is a projection over the entire volume.
% While lower mass black holes are more abundant at overlap, they also
% create fluctuations on smaller scales than larger black holes.  
% Because the heating rate is proportional to mass and we assume a
% random spatial distribution, the two length scales in the system are the
% mean separation of sources, $r_{\rm sep}\propto n_{\rm BH}^{-1/3}$,
% and the distance out to which an individual black hole can effectively
% heat the IGM above the CMB temperature, $r_{\rm heat}\propto M_{\rm BH}^{1/3}$.
% For this simplified case, in which the spatial distribution of black
% holes is random and all black holes have the same mass, the shape of
% the power spectrum is determined only by the ratio $r_{\rm
%   heat}/r_{\rm sep}\propto (n_{\rm BH}M_{\rm BH})^{1/3}=\rho_{\rm BH}^{1/3}$.
% Our choice of $\rho_{\rm BH}=2\times 10^9\ M_\odot{\rm Gpc}^{-3}$ is
% roughly that density at which the signal is maximized; a lower black
% hole mass density results in a lower overall signal, while a higher
% density leads to overlap of the individual regions, and the
% fluctuations are saturated.  

\pagebreak

\subsection{Black hole abundance at high-$z$}

How do the black hole densities and masses we assume here compare to
those observed at $z\sim 6$ and expected at higher redshifts?
\citet{willott/etal:2010}  constructed the mass function of black
holes in the range $10^8\ M_\odot < M_{\rm BH} < 3\times 10^9\ M_\odot$
at $z=6$, finding it to be well-fitted by $dn/d{\rm ln}M_{\rm
  BH}\simeq \phi_*(M_{\rm BH}/M_*)^{-1}\exp(-M_{\rm BH}/M_*)$, with
$\phi_*=5.34~{\rm Gpc}^{-3}$ and $M_*=2.2\times 10^9\ M_\odot$.
Integrating the black hole mass function, one finds $\rho_{\rm
  BH}(>M_{\rm BH})\simeq 7\times 10^9$ and $3\times 10^{10}\
M_\odot{\rm Gpc}^{-3}$, for $M_{\rm BH}=10^9$, and $10^8\ M_\odot$,
respectively, somewhat larger than the black hole mass density we find
which maximizes the 21-cm power spectrum (\S3.2).   Matching the abundance of black holes greater than a given
mass to the dark matter halo mass function of \citet{warren/etal:2006}
at $z=6$ \citep[``abundance matching'' -- e.g.,][]{kravtsov/etal:2004}, we obtain $M_{\rm
    halo}=2.3\times 10^{12}$ and $4.7\times 10^{12}\ M_\odot$ for
  $M_{\rm BH}=10^8$ and $10^9\ M_\odot$, respectively, implying a
  value of $M_{\rm BH}/M_{\rm halo}\simeq 4\times 10^{-5}$ to $2\times
  10^{-4}$ over the same range.  

 More detailed predictions would
require extrapolating the $M_{\rm BH}-M_{\rm halo}$ relationship to
lower masses and higher redshifts, or making highly uncertain
assumptions about the formation mechanism of the high-redshift seeds
and their accretion history.  For example, the
ratio $M_{\rm BH}/M_{\rm halo}\simeq 10^{-4}$ we determine here by
abundance matching at $z=6$ and $M_{\rm BH}=10^8-10^9\
M_\odot$ would be a significant underestimate in atomic cooling halos
where black hole formation by direct collapse took place, in which it
is possible that $M_{\rm BH}/M_{\rm halo}$ could approach the limiting
value of $\Omega_b/\Omega_m\simeq 0.17$.  Clearly much more work is
required in understanding the high-redshift quasar population, and for
this reason the constraints provided by either detection or
non-detection of the signal we predict here would be very valuable.  

% \begin{figure}
% \includegraphics[width=0.44\textwidth]{fig2.pdf}
% \caption{Brightness temperature fluctuations at $z=15$ in a region with a mean
%   black hole density of 1~$M_\odot/$Mpc$^3$.  Shown is a slice through
%   a box $(M_{\rm BH}/10^5~M_\odot)^{1/3}$~Gpc on a side.  For example,
%   this box shows the fluctuations from $10^5$~$M_\odot$ black holes
%   with a number density of $10^4$~Gpc$^{-3}$ in a box 1~Gpc on a side,
%   or equivalently the fluctuations from $10^7$~$M_\odot$ black holes
%   with a number density of 200~Gpc$^{-3}$ in a box about 9~Gpc on a
%   side.  
% \\
% }
% \label{fig2}
% \end{figure}

\begin{figure}
\includegraphics[width=0.49\textwidth]{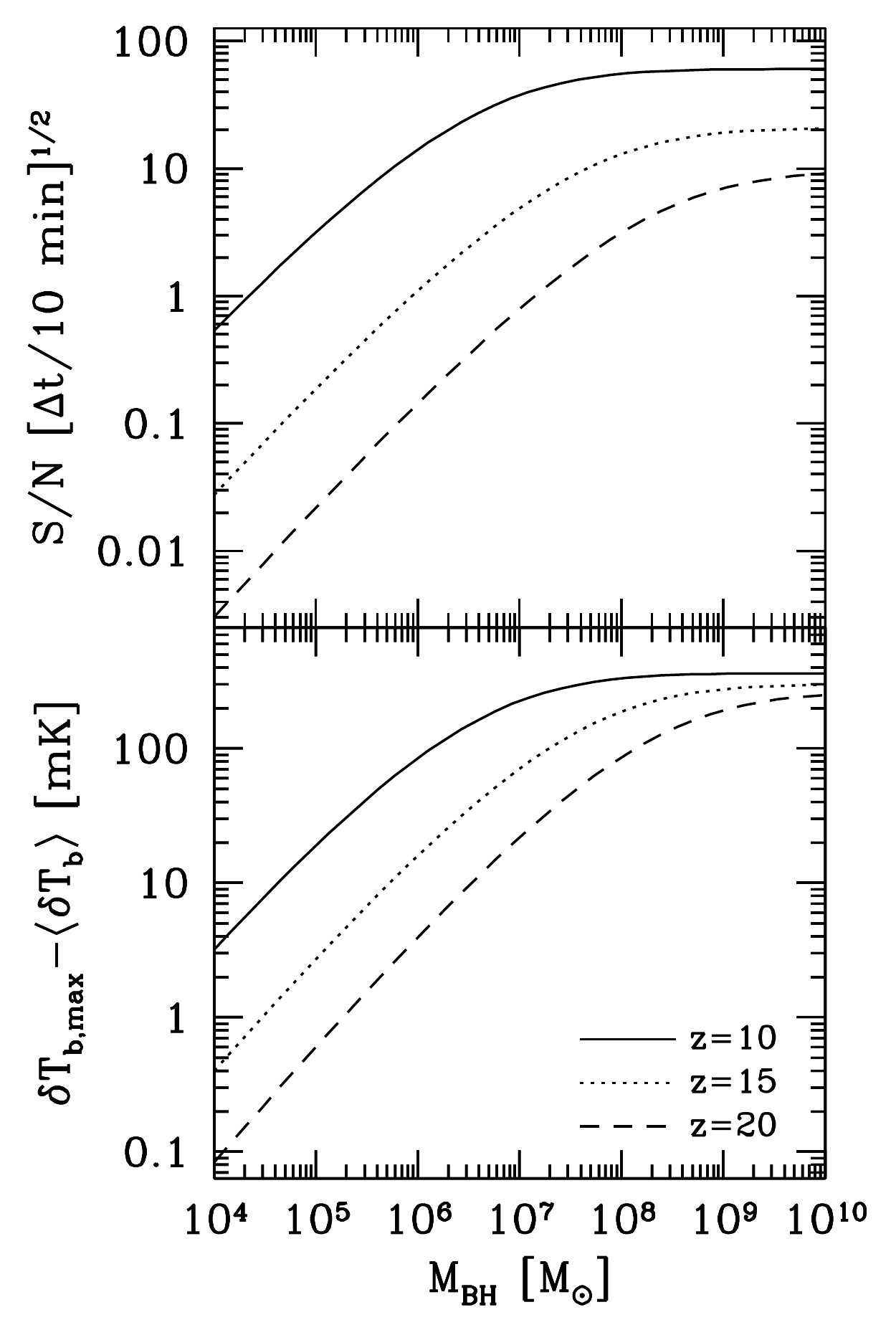}
\caption{ 
Bottom: Peak fluctuation amplitude, measured with respect to the mean
background absorption $\langle \Delta T_{\rm b}\rangle(z)$, as a
function of black hole mass, for three different redshift.
Top: Signal-to-noise ratio at the same redshifts for a bandwidth corresponding to
the beam width and an integration time of 10 minutes.  The
detectability declines rapidly in the interval $10<z<20$.}
\label{fig2}
\end{figure}

\section{Detectability}

In this section, we estimate the detectability of individual sources
as well the statistical detection of their power spectrum.  In the
case of individual objects, we focus on a novel approach, using
single-dish filled aperture telescopes like
Arecibo\footnote{\htmladdnormallink{http://www.naic.edu}{http://www.naic.edu}}
and   
FAST\footnote{\htmladdnormallink{http://fast.bao.ac.cn}{http://fast.bao.ac.cn}},   
while for the
power spectrum we will simply refer to existing sensitivity estimates
for facilities such as
LOFAR\footnote{\htmladdnormallink{http://www.lofar.org}{http://www.lofar.org}}, 
MWA\footnote{\htmladdnormallink{http://www.mwatelescope.org}{http://www.mwatelescope.org}}, 
and 
SKA\footnote{\htmladdnormallink{http://www.skatelescope.org}{http://www.skatelescope.org}}.

\subsection{Individual quasars}

\begin{figure}
\includegraphics[width=0.48\textwidth]{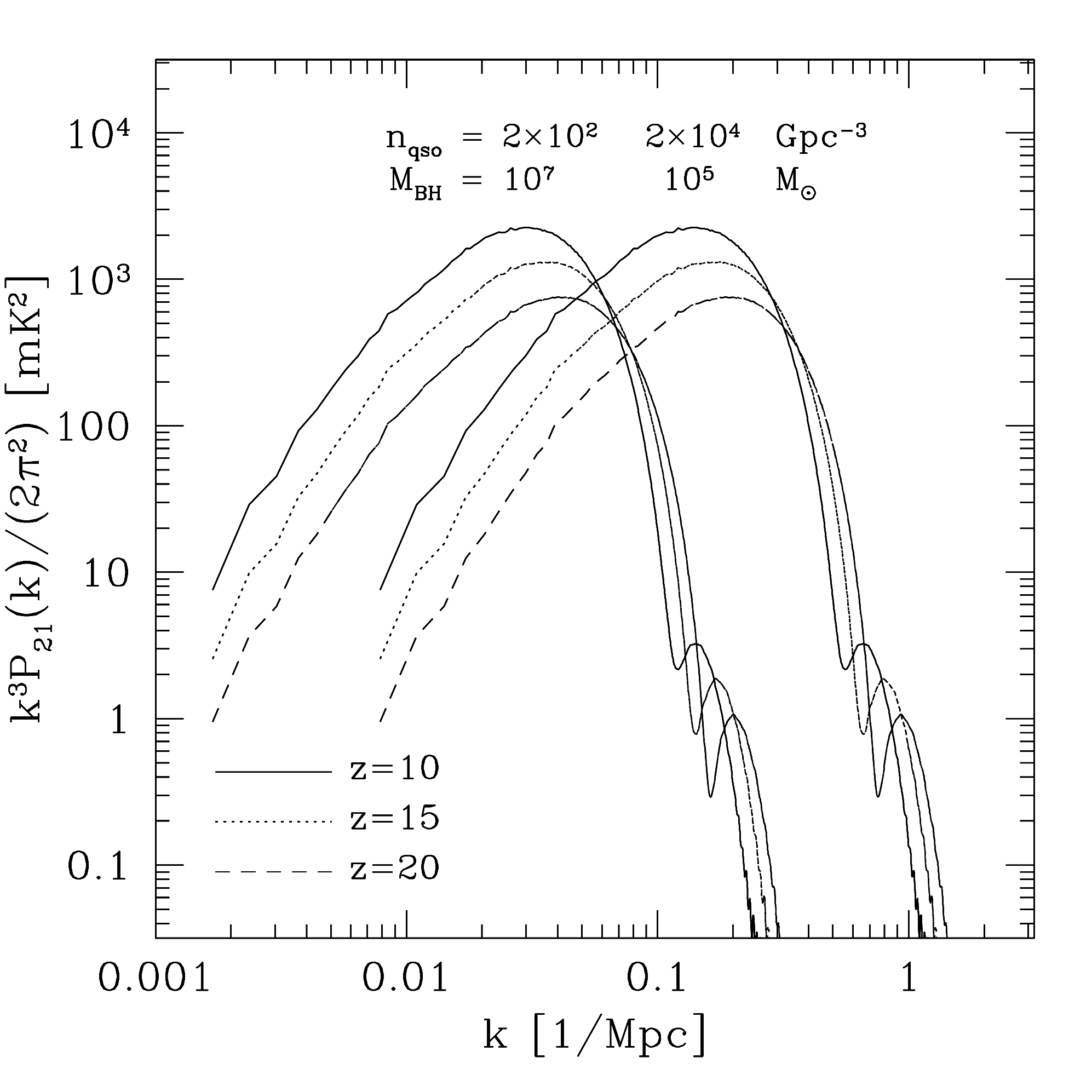}
\caption{ 
Shown is the spherically-averaged three-dimensional power spectrum of
21-cm fluctuations for a black hole density 1~$M_\odot/$Mpc$^3$, for
three different redshifts, as labeled.   Large, relatively rare
$10^7~M_\odot$ black holes have a power spectrum which peaks at $k\sim
0.03$~Mpc$^{-1}$, while for $M_{\rm BH}=10^5~M_\odot$ the power
spectrum which peaks at $k\sim 0.15$Mpc$^{-3}$.  Such a signal at
$z=10$ should be easily detectable by LOFAR, MWA, or SKA.}   
\label{fig3}
\end{figure} 

In order to determine the necessary integration time, we convolve the
profiles shown in Fig.~\ref{fig1} with a half-power beam width of
\beq
\theta_{\rm b}=
26'~\left(\frac{1+z}{11}\right)\left(\frac{d_{\rm dish}}{300~{\rm m}}\right)^{-1},
\label{beam}
\eeq
where $d_{\rm dish}$ is the effective dish diameter.  Converting angle on the sky
to comoving distance, we obtain the comoving resolution,
\beq
D\simeq 70~{\rm Mpc}\left(\frac{d_{\rm dish}}{300~{\rm m}}\right)^{-1}
\left(\frac{1+z}{11}\right)^{1.2}.
\eeq
This implies that the fluctuation created by a $10^7 M_\odot$-black
hole would be just resolvable with a single 300-m dish like Arecibo
(see Fig.~\ref{fig1}), lower mass black holes would be unresolved
point sources, and the profile of higher mass black holes could
actually be measured. For an integration time of $\Delta t$, the
sensitivity is given by  
\beq
\delta T_{\rm err}=22~{\rm mK}
\left(\frac{\Delta t}{60~{\rm s}}\right)^{-1/2}
\left(\frac{d_{\rm dish}}{300~{\rm m}}\right)^{1/2}
\left(\frac{1+z}{11}\right)^{2.35}
\eeq
where we have used 
\beq
\Delta\nu = 4~{\rm MHz}
\left(\frac{D}{70~{\rm Mpc}}\right)
\left(\frac{1+z}{11}\right)^{-1/2}
\eeq
for the bandwidth corresponding to a comoving distance $D$,
$\delta T_{\rm err}=T_{\rm sys}/\sqrt{\Delta\nu\Delta t}$, and $T_{\rm
  sys}\simeq 3\times 10^5$~mK$[(1+z)/11]^{2.7}$ 
\citep[e.g.,][]{furlanetto/etal:2006}.   

We calculate the maximum fluctuation amplitude as a function of quasar black
hole mass and redshift, $\delta T_{\rm b,max}(M_{\rm  bh},z)$,
which is a convolution of the beam with the individual profiles
plotted in Fig.~\ref{fig1}, 
\beq
\delta T_{\rm b,max}=\frac{2}{\theta_{\rm
   g}^2D}
\int_0^{\infty} d\theta
\theta e^{-\frac{\theta^2}{2\theta_{\rm g}^2}}
\int_0^{D/2}dl
\delta T_b(r_{l\theta}),
\eeq
where we use a Gaussian profile with a FWHM of $\theta_{\rm b}(z)$,
such that $\theta_{\rm b}(z)=2\sqrt{2\ln 2}\theta_{\rm g}(z)$, and
$r_{l\theta}^2=r_\theta^2+l^2$, where $r_\theta$ is the projected
comoving distance perpendicular to the line of sight corresponding to
the angle $\theta$.

Shown in Fig.~\ref{fig2} are the resulting fluctuation amplitudes with
respect to the mean back ground absorption, 
$\delta T_{\rm
 b,max}(M_{\rm bh},z)-\langle\delta T_b\rangle (z)$, as well as the 
signal-to-noise ratio, 
$[\delta T_{\rm b,max}(M_{\rm bh},z)-\langle\delta T_b\rangle
(z)]/\delta T_{\rm b,err}(z)$ for an integration time of 10 minutes.
As can be seen from the figure both the signal, and to a greater
extent the signal-to-noise, decline rapidly with increasing redshift.
This is due to several reasons.  First, the intrinsic signal declines
with increasing redshift because absorption is relatively weaker at
higher redshifts when the IGM has not had as much time to cool (see Fig.~\ref{fig1}).  In
addition, the beamwidth gets larger, in proportion to $1+z$, while the
angular size of the fluctuations at fixed black hole mass actually
decline with redshift, as seen from Fig.~\ref{fig1}.

\subsection{Power spectrum from Poisson fluctuations}

To estimate the fluctuating background from a superposition of sources, we
assume all black holes at a given redshift have the same mass, $M_{\rm
  BH}$,  and comoving number density, $n_{\rm BH}$.  We then
generate a realization of random black hole positions within a
periodic box, and calculate the heating from each source using
equation (\ref{ts}).  Because the heating rate is proportional to mass and we assume a
random spatial distribution, the two length scales in the system are the
mean separation of sources, $r_{\rm sep}\propto n_{\rm BH}^{-1/3}$,
and the distance out to which an individual black hole can effectively
heat the IGM above the CMB temperature, $r_{\rm heat}\propto M_{\rm BH}^{1/3}$.
For this simplified case, in which the spatial distribution of black
holes is random and all black holes have the same mass, the shape of
the power spectrum is determined only by the ratio $r_{\rm
  heat}/r_{\rm sep}\propto (n_{\rm BH}M_{\rm BH})^{1/3}=\rho_{\rm BH}^{1/3}$.
Our choice of $\rho_{\rm BH}=2\times 10^9\ M_\odot{\rm Gpc}^{-3}$ is
roughly that density at which the signal is maximized; a lower black
hole mass density results in a lower overall signal, while a higher
density leads to overlap of the individual regions, and the
fluctuations become saturated.   

Shown in Fig. \ref{fig3} is our predicted power spectrum for Poisson
fluctuations of black holes of mass $M_{\rm BH}=10^5$
and $10^7 M_\odot$ at $z=$10, 15, and 20.  Because the black hole mass
density is the same in both cases, the curves have the same shape, but are shifted in
wavenumber such that $k^{-3}\propto M_{\rm BH}$.  Because the
individual regions are only weakly overlapping for $\rho_{\rm BH}=2
\times 10^9\ M_\odot{\rm Gpc}^{-3}$, the shape of the curve is
quite close to the Fourier transform of an individual region, so that
\beq
P(k)^{1/2}\propto 
\int_0^\infty r^2dr \delta T_b(r)\frac{\sin kr}{kr},
\eeq
with $\delta T_b(r)$ given by eqs. (\ref{dtb}) and (\ref{ts}).  

The amplitude of this signal at $k\sim 0.1\ {\rm Mpc}^{-1}$
($[k^3P(k)/(2\pi^2)]^{1/2}\simeq 50\ {\rm mK}$) 
is almost an order of magnitude greater than
that expected from ionized bubbles when $T_S\gg T_{\rm CMB}$ 
\citep[$\simeq 6\ {\rm mK}$, e.g.,][]{furlanetto/etal:2004}, which 
compensates for the increased foregrounds at the higher redshifts
corresponding to the signal we predict here.  For example,
\citet{mcquinn/etal:2006} estimated the sensitivity of various
facilities to the spherically-averaged power spectrum, finding (at $z=12$) $\delta
T^2_{\rm b,err}(k\sim 0.1 {\rm Mpc}^{-1})\simeq 30 {\rm mK}^2$  
for LOFAR and MWA, and  $T^2_{\rm b,err}(k\sim 0.1 {\rm
  Mpc}^{-1})\simeq 0.1 {\rm mK}^2$ for SKA, for their adopted array
configurations and 1000 hr of integration time \citep[Fig. 6 and Table
1 of][]{mcquinn/etal:2006}.  Thus, the power spectrum shown in Fig.~4
would be easily detectable by either of these three experiments at
$z\sim 12$ for the survey parameters used by \citet{mcquinn/etal:2006}. 

%\subsection{Observational Strategies for Arecibo and FAST}

\section{Survey strategies}

% Using a simple parametrization of the abundance and luminosities of
% the first quasars, we have calculated their detectability over a broad
% range of possible black hole masses and densities, from $M_{\rm
%   BH}=10^5 M_\odot$ and $n_{\rm BH}\simeq 2\times 10^4$~Gpc$^{-3}$, 
% to $M_{\rm BH}=10^9 M_{\odot}$ and $n_{\rm BH}\simeq 2$~Gpc$^{-3}$,
% over the redshift range $10<z<20$. 

Since the signal comes from a large angular scale on the sky, a filled
aperture maximizes the sensitivity.  The two largest collecting area
telescopes are Arecibo and, under 
construction, FAST.  They are at latitudes $\delta=18$ (26) degrees,
respectively.  To stabilize the baselines, a drift scan at 80 MHz maps
a strip of width 40 (90) arc minutes by length 360 degrees
$\cos{\delta}$ every day.  With Arecibo, a custom feedline with
pairwise correlations of dipoles would allow a frequency dependent
illumination of the mirror, allowing the frequency independent removal
of foregrounds.  It would also allow operation as an interferometer,
increasing stability and rejection of interference, and enabling
arbitrary apodization of the surface.  Should grating
sidelobes from support structures become a problem, one could also
remove the carriage house, and mount the feedline on a pole from the
center of the dish.  This would result in a clean, unblocked aperture
with frequency independent beam.  Even with the carriage support
blocking, the side lobes would still be frequency independent for an
appropriately scaled illumination pattern.

For FAST, a focal plane array also enables a frequency dependent
illumination of the primary, resulting in a frequency independent
beam on the sky.  It also increases the survey speed by the number of
receivers used.  Only one receiver is needed every half wave length,
roughly two meters.  Hundred pixel surveys seem conceivable at low
cost, since only small bandwidths would be needed, and the system
temperature is sky limited, even with cheap TV amplifiers.

\pagebreak

We note that the ratio of signal to foreground in this regime is
comparable to that during reionization.  With a filled
aperture, it may be easier to achieve foreground subtraction.  This has been
demonstrated for intensity mapping at $z\sim 0.8$ using the filled
aperture of GBT \citep{chang/etal:2010}, where a similar
foreground ratio exists of galactic synchrotron to 21cm. 

\begin{figure}
\includegraphics[width=0.48\textwidth]{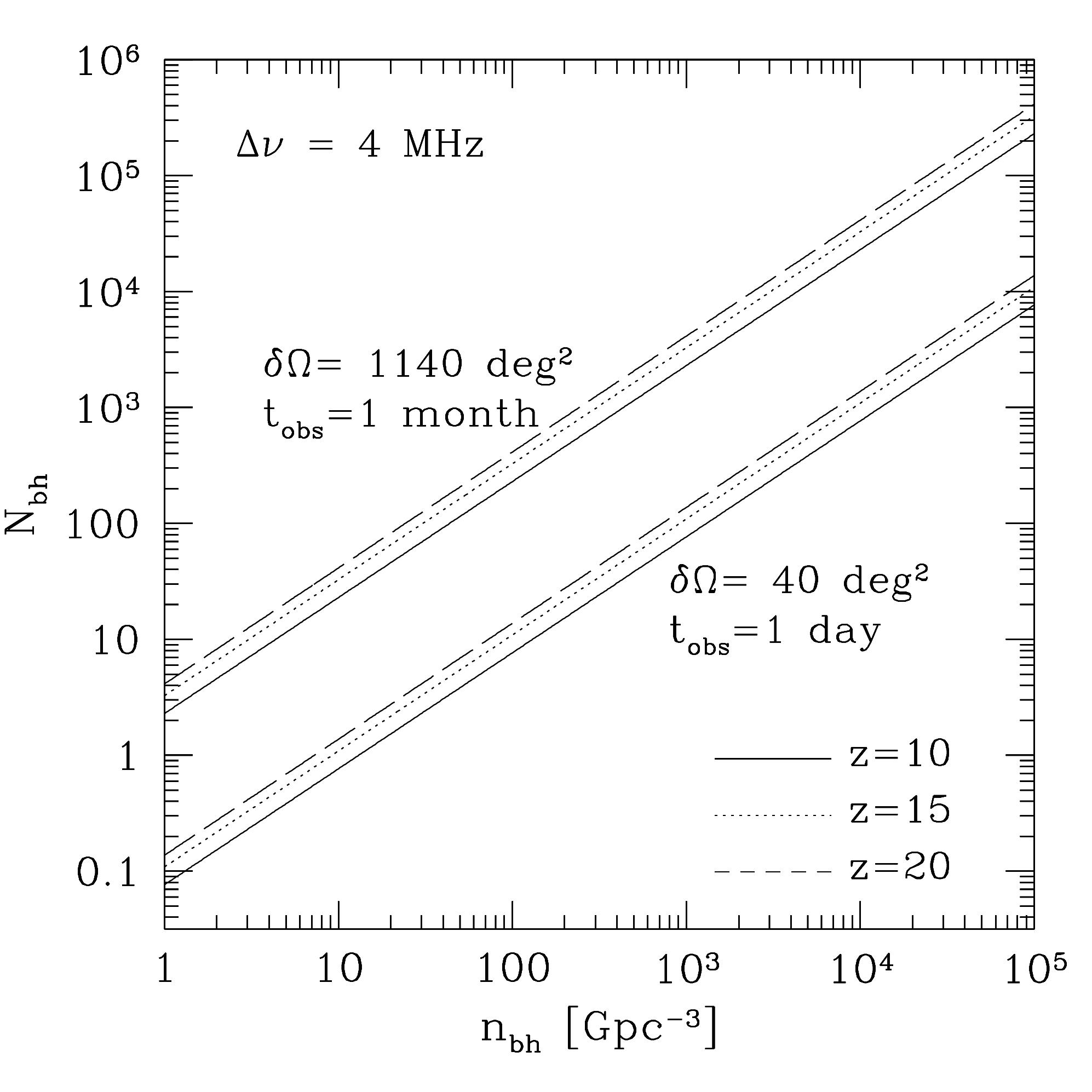}
\caption{Number of black holes expected within a 1-month (upper
curves) and 1-day (lower curves) survey in drift-scan 
mode that scans each point in the sky nine times, as a function of 
comoving black hole density.  The integration time 
per pixel in each case is about 7, 14, and 25 minutes for $z=10$, 15, 
and 20.  With a 14-minute integration time, a $10^7-M_\odot$ black hole
would be detectable at greater than 5-$\sigma$ significance (see Fig.~\ref{fig2})}
\label{fig4}
\end{figure} 

Shown in Fig.~\ref{fig4} is the expected number of black holes within
the field of view for a given redshift and survey bandwidth, versus
their comoving number density.  A dedicated one month survey operating
at 88~MHz ($z\sim 15$) with a 300-m dish like Arecibo, which scans
each point in the sky nine times
would have a pixel size of about 0.4~deg$^2$ during which each pixel
was integrated for about 15 minutes and 1140~deg$^2$ were surveyed.
If $10^7-M_\odot$ black holes had a number density of 1~Gpc$^{-3}$ at
that time, one would expect to discover about three or four of them at
greater than 5-$\sigma$ significance since each pixel would have been
integrated for about 15 minutes at $z=15$ (see dotted lines in
Figs. \ref{fig2} and \ref{fig4}).  Black holes with masses greater
than $10^7~M_\odot$ and similar abundances would be easily detectable,
allowing for followup with longer baseline facilities such as
LOFAR to determine the detailed shape of their 21-cm profile.  
Detecting a $10^6-M_\odot$ black hole at 5-$\sigma$ significance would require a
significantly longer integration time on each pixel, about 4 hours,
so that only about 70~deg$^2$ could be surveyed in a month.  Thus, the
minimum spatial density of $10^6-M_\odot$ black holes would be about
5~Gpc$^{-3}$ at $z=15$.  Smaller black holes would mean even smaller
fields of view and longer integration times per pixel, so in those
cases detecting their statistical signature in the
spherically-averaged power spectrum discussed in \S3.2 may be a better
approach. 

Regardless of whether high-redshift quasar X-ray sources will be
detected directly by the means we propose here, it is clear that the
pre-reionization universe offers a wealth of information that can be
uniquely probed with the 21-cm transition, and it is therefore vital that
theoretical models for the signal originating from before the epoch of
reionization continue to be developed.

% If quasars hosting $10^9 M_\odot$ black holes existed before the mean
% IGM was heated above the CMB temperature, we find that an observatory
% like Arecibo could detect and image their corresponding partially-heated
% halos at the 5-$\sigma$ level by surveying a substantial fraction of
% the sky within a few months at a frequency of 90 MHz ($z\sim 15$) with
% a bandwidth of $\Delta\nu=4$~MHz.   If the brightest quasars at the
% end of the heating phase instead host black holes with a mass of $10^5
% M_\odot$, then their Poisson fluctuations could be detected at
% 3-$\sigma$ significance over a year's time.

\vspace{1cm}

\acknowledgments{We wish to thank J.~Peterson and R.~M.~Thomas for
  helpful discussions on Arecibo and 21-cm signatures of early QSOs,
  resepectively, and J.~R.~Pritchard for comments on an earlier draft of
  the paper.  M.~A.~A. and T.~C. are grateful for the hospitality of
  the  Aspen Center for Physics, where this work was completed.  We
  acknowledge financial support by CIfAR and NSERC.}

\bibliography{references}
\bibliographystyle{apj}

\end{document}